\documentclass{ws-procs975x65}
\input{epsf}

\begin{document}

\title{\large Solution for "geodesic" motion of a Schwarzschild black hole\\
along a magnetic field in AdS${}^2\times \mathbb{S}^2$ space-time}

\author{George A. Alekseev}

\address{Steklov Mathematical Institute of Russian Academy of Sciences,\\
 Gubkina 8, 119991, Moscow Russia\\
E-mail: G.A.Alekseev@mi.ras.ru}

\begin{abstract}
The exact solution of Einstein - Maxwell equations for a Schwarzschild black hole immersed in the static spatially homogeneous  AdS${}^2\times\mathbb{S}^2$ space-time of  Bertotti-Robinson magnetic universe is presented. In this solution, the black hole possesses a finite  initial boost in the direction of the magnetic field and performs a ``geodesic'' oscillating motion interacting with the background gravitational and electromagnetic  fields.

\end{abstract}


\bodymatter

\section*{Introduction}
The century which passed since a wonderful discovery of General Relativity, brought us a lot of developed methods for construction of exact solutions of Einstein and Eisntein-Maxwell field equations and findings of numerous particular solutions \cite{SKMHH:2003, Griffiths-Podolsky:2009, Alekseev:2011}. However, despite the permanent interest in the literature to different aspects of  solutions which describe an interaction of various compact sources (such as black holes, particle-like singularities, matter discs, rings, etc) with the external gravitational and electromagnetic fields, very few such exact solutions are known which describe some non-trivial motion of the sources in various external fields.

In this paper, we present a construction of new solution of Einstein-Maxwell equations for a Schwarzschild black hole moving freely along a magnetic field in  AdS${}^2\times \mathbb{S}^2$ space-time of Bertotti-Robinson magnetic universe shown on Fig. 1.
\vspace{-6ex}
\begin{figure}[h]
\begin{center}
\epsfig{file=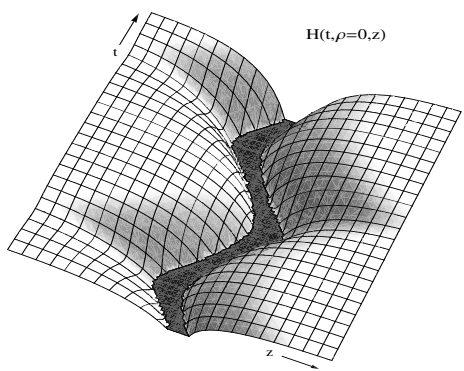,width=2.8in,height=2.3in}
\vspace{-2ex}
\caption{Schwarzschild black hole in a "geodesic" motion along a magnetic field in Bertotti-Robinson (BR) magnetic universe. In Weyl-like coordinates, the black hole horizon is represented by segments on the axis $\rho=0$
which constitute a black stripe on $(t,z)$-plane. $H=g_{\scriptscriptstyle{\mathcal{TT}}}/ g^{(BR)}_{\scriptscriptstyle{\mathcal{TT}}}$ and  $g_{\scriptscriptstyle{\mathcal{TT}}}$ is the square of time-like Killing vector field $\partial/\partial\mathcal{T}$ which becomes null on the horizon.}
\end{center}
\end{figure}

Among the earlier found solutions which admit some dynamical interpretation it is worth to mention the Ernst solution in Ref.~\refcite{Ernst:1976} which describes a charged black hole immersed in Melvin electric universe and accelerated by its electric field (see also Ref.~\refcite{Bicak-Kofron:2010} for details and references) and solution derived in Ref.~\refcite{Alekseev-Garcia:1996} for a Schwarzschild black hole immersed in Bertotti-Robinson magnetic universe and resting at the origin of some rigid frame without any struts or string-like singularities.

In comparison with the solution Ref.~\refcite{Alekseev-Garcia:1996}{}, the solution presented here possess one more arbitrary real parameter which determines the initial boost of a black hole along the direction of the magnetic field. This boosted black hole performs a free "geodesic" motion, oscillating around the origin of a chosen rigid reference frame.
Below, we consider at first the motion of a test particles along the magnetic field in the
Bertotti-Robinson background and show that any such particle performs an oscillating
motion around the origin of some rigid reference frame where a geodesic particle with zero boost can be at rest. Then
 we show that a similar rigid frame can be associated with any other geodesic world line
directed along the magnetic field. In each of these frames the metric in co-moving coordinates
coincides with the same Bertotti-Robinson one. Therefore, we can use this metric as the background and apply the solution generating methods developed earlier for Einstein-Maxwell equations\footnote{See, in particular, Ref.~\refcite{Alekseev:1980} for generation of Einstein-Maxwell solitons or Ref.~\refcite{Alekseev:1985, Alekseev:1988} for application of the monodromy transform approach and the corresponding integral equation method.} for construction of the solution with a black hole immersed in this metric, that was done in Ref.~\refcite{Alekseev-Garcia:1996}.
The generated solution is static in a co-moving frame, but in the original frame it looks like a free oscillating motion of a black hole near its equilibrium position without struts and string-like singularities on the axis.

\section*{The background Bertotti-Robinson space-time with AdS${}^2\times \mathbb{S}^2$ topology and spatially homogeneous magnetic field}
The space-time of the Bertotti-Robinson magnetic universe possess AdS${}^2\times \mathbb{S}^2$ topology and it is filled by spatially homogeneous magnetic field (see Ref.~\refcite{SKMHH:2003, Griffiths-Podolsky:2009} for details and references). In cylindrical coordinates $\{t,\rho,z,\varphi\}$, the components of its metric, 1-form of electromagnetic potential and the only non-zero component of magnetic field measured by the local observer can be presented in the form ($\gamma=c=1$):
\begin{equation}\label{BR-solution}
\begin{array}{l}
ds^2=-\cosh^2\! \left(\dfrac zb\right)\,dt^2+d\rho^2+dz^2+b^2
\sin^2\!\left(\dfrac{\rho}{b}\right)\,d\varphi^2\\[2ex]
\underline{{\mathbf{A}}}=\left\{0, 0,0,-2 b \sin^2\!\left(\dfrac {\rho}{2 b}\right)\right\},\qquad
H_{\widehat{z}}=\dfrac 1b
\end{array}
\end{equation}
where $b$ is a constant parameter. The sections $\{t=const,z=const\}$ possess the geometry of a 2-sphere of the radius $b$ and the space-time is closed in the $\rho$-direction.
In the limit $b\to\infty$, the magnetic field vanishes ($H_{\widehat{z}}\to
0$) and the metric (\ref{BR-solution}) transforms into the usual Minkowski
metric in cylindrical coordinates.

\section*{Geodesic motion of test particles along the $z$-axis in the metric  (\ref{BR-solution})}
The world lines of test particles which are at rest in the
metric (\ref{BR-solution}) constitute a rigid frame. These world lines possess the acceleration along $z$-axis which value is
\[W^z=\dfrac{1}{b}\tanh\dfrac{z}{b}
\]
and only the world line of a test particle at rest in the position  $z=0$ is geodesic.
The time-like geodesics which correspond to a motion of test particles of mass $\mu$ and the energy $E_o$ along $z$-axis in metric (\ref{BR-solution}) are
\begin{equation}\label{BR-geodesics}
\tanh\dfrac zb=\dfrac{\sqrt{\mathcal{E}_o^2-1}}{\mathcal{E}_o}\sin\dfrac{t-t_{\ast}}{b}
,\qquad \qquad
\mathcal{E}_o=\dfrac{E_o}{\mu}>1
\end{equation}
This equation shows that the motion of a freely falling neutral test particle boosted with some finite energy $\mathcal{E}_o > 1$ along the $z$-axis in the metric (\ref{BR-solution}) is periodic and it takes place in some restricted region $\cosh\dfrac {z}{b}<\mathcal{E}_o$.
\begin{figure}
\begin{center}
\parbox{2.1in}{\epsfig{figure=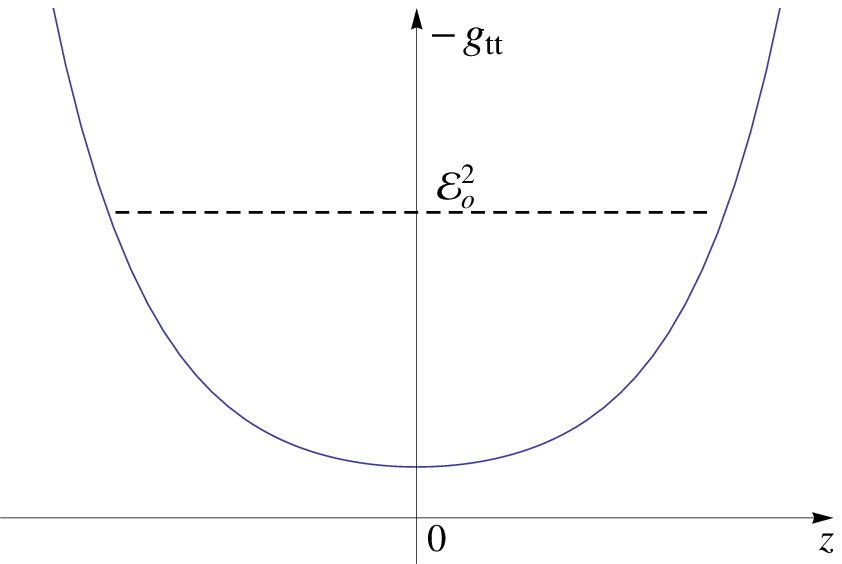,width=2.1in}
}
\hspace*{10pt}
\parbox{2.1in}{\epsfig{figure=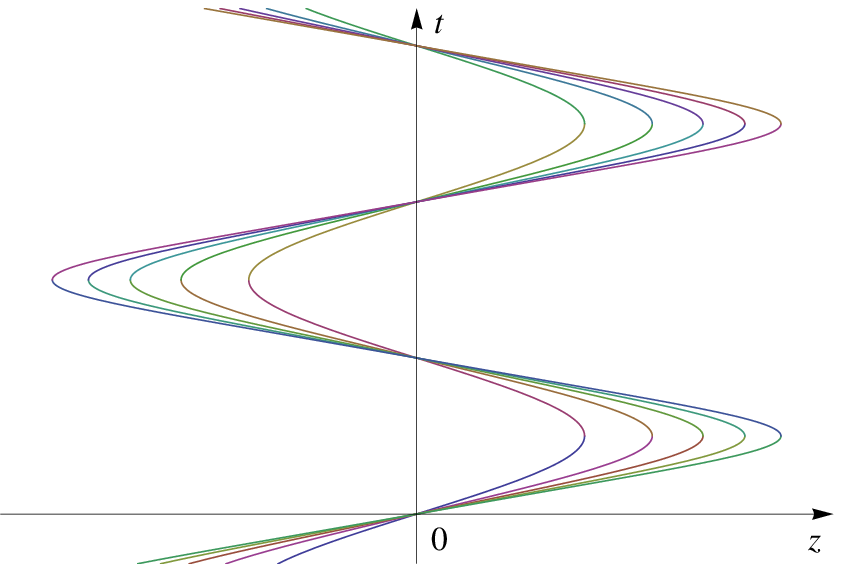,width=2.1in}
}
\caption{
The left picture shows the form of $g_{tt}$ metric component which plays the role of a potential well for test particles moving with the energy $\mathcal{E}_o$ along a magnetic field in Bertotti-Robinson magnetic universe. The right picture shows the geodesic world lines for these test particles on $(t,z)$ plane for different energies $\mathcal{E}_o$.}
\end{center}
\end{figure}

\section*{Killing vector fields in $(t,z)$-plane}
The rigid frames in the metric (\ref{BR-solution}) are determined by congruences of Killing vector fields in $(t,z)$-plane. In the coordinates $(t,\rho,z,\varphi)$ these can be expressed as:
\begin{equation}\label{Killings}
\xi^i=\{\cosh\delta-\sinh\delta\sin\dfrac {t-t_\ast}b\tanh\dfrac z b,\,\, 0,\,
\sinh\delta\cos\dfrac {t-t_\ast}b,\,\,0\}
\end{equation}
where $\delta$ and $t_\ast$ are arbitrary real parameters. Thus, at every point of $(t,z)$-plane we have a family of Killing vector fields parametrized by $\delta$ which possess the norm
\[
g_{\xi\xi}\equiv \xi_i\xi^i=-\cosh^2\dfrac zb\left(\cosh\delta-\sinh\delta\,\sin\dfrac {t-t_\ast}b\tanh\dfrac zb\right)^2+\sinh^2\delta\,\cos^2\dfrac {t-t_\ast}b .
\]

\section*{Rigid frames associated with geodesics on $(t,z)$-plane}
The background solution (\ref{BR-solution}) possess an important symmetry: changing the reference frame with corresponding coordinate transformation  $(t,\rho,z,\varphi)\to (\mathcal{T},\rho,\mathcal{Z},\varphi)$ with
\begin{equation}\label{TZ}
\left\{\begin{array}{l}
\mathcal{T}=b\,\arctan\!\left[\cosh\delta\,\,\tan\!\dfrac {t-t_\ast}b-\sinh\delta\,\tanh\dfrac{z}{b}\,\, \hbox{sec}\!\dfrac{t-t_\ast}{b}
\right],\\[2ex]
\mathcal{Z}=b\,\hbox{arcsinh}\left[\cosh\delta\,\sinh\dfrac zb-\sinh\delta\,\cosh\dfrac zb\,\,\sin\dfrac {t-t_\ast}b\right],
\end{array}
\right.
\end{equation}
leavs the form of the background metric invariant:
\[
ds^2=-\cosh^2\!\dfrac zb\,\,\, dt^2+dz^2+d\rho^2+\sin^2\dfrac{\rho}b d\varphi^2=-\cosh^2\!\dfrac {\mathcal{Z}}b\,\, d\mathcal{T}^2+d{\mathcal{Z}}^2+d\rho^2+\sin^2\dfrac{\rho}b d\varphi^2
\]
Thus, we constructed a rigid frame which motion with respect to the original rigid frame is determined by the boost parameter $\delta$. Among the world lines ${\mathcal{Z}}=const$  only the line ${\mathcal{Z}}=0$ is geodesic and for this we have
\[\mathcal{E}_o=\cosh\delta
\]
The other Killing lines with ${\mathcal{Z}}=const\ne 0$ and with the same $\delta$ are not geodesics.

\section*{Schwarzschild black hole in a "geodesic" motion in metric (\ref{BR-solution})}
The solution for a Schwarzschild black hole at rest in the external Bertotti-Robinson magnetic universe was constructed in Ref.~\refcite{Alekseev-Garcia:1996}\footnote{It is necessary to mention here that, in contrast to such "local" (in the region near the axis $\rho=0$) interpretation of this solution as a black hole in the external field, its global interpretation is more difficult because the semi-closed AdS${}^2\times \mathbb{S}^2$ topology of this external space-time and focusing properties of gravitational field give rise to existence of naked singularity located on the "antipodal" axis $\rho=\pi b$  (see  Ref.~\refcite{Alekseev-Garcia:1996} for more details).}. In a slightly different notations it takes the form \vspace{-3ex}
\[
ds^2=-H \left(\cosh\dfrac{\mathcal{Z}}{b}\right)^2\, d\mathcal{T}^2+f(d\rho^2+d\mathcal{Z}^2)+\dfrac{b^2
\left(\sin\dfrac{\rho}{b}\right)^2}{H}\, d\varphi^2
\]
where the functions $H$, $f$ and the electromagnetic potential $\mathbf{A}=\{0,0,0,A_{\varphi}\}$ are
\[
H=\dfrac{(x_2+y_1)^2}{x_2^2-1},\hskip1ex
A_{\varphi}=-\dfrac {b(x_2+1)(1+y_1)}{x_2+y_1},\hskip1ex f=\dfrac{(x_2+y_1)^2}{x_2^2-y_2^2}
\left[\dfrac{x_1+x_2-y_1-y_2}
{x_1+x_2+y_1+y_2}\right]^2
\]
and the bipolar coordinates $\{x_1,y_1\}$, $\{x_2,y_2\}$ are functions of $\rho$ and $\mathcal{Z}$ only:
\[\left\{\begin{array}{l}
x_1=-\dfrac{b}{m}\sinh\dfrac{\mathcal{Z}}{b},\\[2ex]
y_1=-\cos\dfrac{\rho}{b},
\end{array}
\qquad
\right\{\begin{array}{l}
x_2=(R_1+R_2)/2 m,\quad y_2=(R_1-R_2)/2 m
\\[2ex]
R_{\pm}=\sqrt{(b\sinh\dfrac{\mathcal{Z}}{b}\cos\dfrac{\rho}{b}\mp m)^2+b^2 \cosh^2\dfrac{\mathcal{Z}}{b}\sin^2\dfrac{\rho}{b}}
\end{array}
\]
However, this solution will describe the boosted Schwarzschild black hole
moving in Bertotti-Robinson background, if $\mathcal{T}$ and $\mathcal{Z}$ are functions of $t$ and $z$ defined in (\ref{TZ}). For small mass $m$,  this black hole produces perturbations of metric and electromagnetic field  localized near $\mathcal{Z}=0$ that is a background geodesic oscillating near $z=0$.

\section*{Acknowledgements}
This work was supported in part by the Russian Foundation for Basic Research
(Grants No. 14-01-00049 and No. 14-01-00860).

\end{document}